\documentclass[a4paper,fleqn]{cas-sc}

\usepackage[authoryear, compress]{natbib}

\usepackage{graphicx} 
\usepackage[version=4,arrows=pgf-filled, mathfontname=mathsf]{mhchem} 

\usepackage{siunitx} 
\usepackage{hyperref} 
\usepackage{amsmath} 
\usepackage[utf8]{inputenc}

\def\tsc#1{\csdef{#1}{\textsc{\lowercase{#1}}\xspace}}
\tsc{WGM}
\tsc{QE}

\begin{document}
\let\WriteBookmarks\relax
\def\floatpagepagefraction{1}
\def\textpagefraction{.001}

\shorttitle{Nanoindentation and high-pressure structural behavior of lunar olivine}    

\shortauthors{P. Grèbol-Tomàs et al.}  

\title [mode = title]{Mechanical softening and enhanced elasticity of lunar olivine probed via nanoindentation and high-pressure X-ray diffraction measurements}  



%




\author[1,2]{P. Gr\`{e}bol-Tom\`{a}s}[orcid=0009-0004-2554-890X]

\affiliation[1]{organization={Institut de Ci\`{e}ncies de l'Espai (ICE-CSIC)},
                addressline={C/ Can Magrans, s/n}, 
                city={Cerdanyola del Vall\`{e}s},
                postcode={08193}, 
                state={Barcelona, Catalonia},
                country={Spain}}
\affiliation[2]{organization={Insitut d'Estudis Espacials de Catalunya (IEEC)},
                addressline={C/ Esteve Tarradas 1, Parc Mediterrani de Tecnologia (PMT) Campus Baix Llobregat - UPC}, 
                city={Castelldefels},
                postcode={08860}, 
                state={Barcelona, Catalonia},
                country={Spain}}

\author[3]{J. Ib\'{a}\~nez-Insa}[orcid=0000-0002-8909-6541]
\cormark[1]
\ead{jibanez@geo3bcn.csic.es}
\affiliation[3]{organization={Geosciences Barcelona (GEO3BCN-CSIC)},
                addressline={C/ Llu\'{i}s Sol\' {e} i Sabar\'{i}s, s/n}, 
                city={Barcelona},
                postcode={08028}, 
                state={Catalonia},
                country={Spain}}
\author[1,2]{J. M. Trigo-Rodr\'{i}guez}

\author[4]{E. Pe\~na-Asensio}[orcid=0000-0002-7257-2150]
\affiliation[4]{organization={Department of Aerospace Science and Technology, Politecnico di Milano},
                addressline={Via La Masa 34}, 
                city={Milano},
                postcode={20156}, 
                country={Italy}}

\author[3]{R. Oliva}[orcid=0000-0002-9378-4048]

\author[3]{D. D\'{i}az-Anichtchenko}

\author[5]{P. Botella}
\author[5]{J. S\'{a}nchez-Mart\'{i}n}
\author[5]{R. Turnbull}[orcid=0000-0001-7912-0248]
\author[5]{D. Errandonea}
\affiliation[5]{organization={Departamento de F\'{i}sica Aplicada-ICMUV, MALTA Consolider Team, Universidad de Valencia},
                addressline={C/ Dr. Moliner 50}, 
                city={Burjassot},
                postcode={46100}, 
                state={Valencia},
                country={Spain}}

\author[6]{A. Liang}
\affiliation[6]{organization={Centre for Science at Extreme Conditions and School of Physics and Astronomy, University of Edinburgh},
                city={Edingurgh},
                postcode={EH9 3FD},
                state={Scotland},
                country={United Kingdom}}

\author[7]{C. Popescu}
\affiliation[7]{organization={CELLS-ALBA Synchrotron Light Facility},
                addressline={C/ de la Llum 2-26}, 
                city={Cerdanyola del Vall\`{e}s},
                postcode={08290}, 
                state={Barcelona, Catalonia},
                country={Spain}}

\author[8, 9]{J. Sort}
\affiliation[8]{organization={Departament de Fı\'{i}sica, Universitat Aut\`{o}noma de Barcelona (UAB)},
                addressline={C/ dels Til·lers s/n}, 
                city={Cerdanyola del Vall\`{e}s},
                postcode={08193}, 
                state={Barcelona, Catalonia},
                country={Spain}}
\affiliation[9]{organization={Instituci\'{o} Catalana de Recerca  Estudis Avançats (ICREA)},
                addressline={Passeig Llu\'{i}s Companys 23}, 
                city={Barcelona},
                postcode={08010}, 
                state={Catalonia},
                country={Spain}}

\cortext[cor1]{Corresponding authors}


\begin{abstract}
The mechanical properties of minerals in planetary materials are not only interesting from a fundamental point of view but also critical to the development of future space missions. Here we present nanoindentation experiments to evaluate the hardness and reduced elastic modulus of olivine, \ce{(Mg, Fe)2SiO4}, in meteorite NWA~12008, a lunar basalt. Our experiments suggest that the olivine grains in this lunaite are softer and more elastic than their terrestrial counterparts. Also, we have performed synchrotron-based high-pressure X-ray diffraction (HP-XRD) measurements to probe the compressibility properties of this meteorite and, for comparison purposes, of three ordinary chondrites. The HP-XRD results suggest that the axial compressibility of the orthorhombic $b$ lattice parameter of olivine relative to terrestrial olivine is higher in NWA~12008 and also in the highly-shocked Chelyabinsk meteorite. The origin of the observed differences is discussed. A simple model combining the results of both our nanoindentation and HP-XRD measurements allows us to describe the contribution of macroscopic and chemical-bond related effects, both of which are necessary to reproduce the observed elastic modulus softening. Such joint analysis of the mechanical and elastic properties of meteorites and returned samples opens up a new avenue for characterizing these highly interesting materials.
\end{abstract}


\begin{highlights}
\item Lunar olivine in meteorite NWA12008 is softer and more elastic than terrestrial olivine, as revealed by nanoindentation experiments.
\item HP-XRD measurements show that the $b$ orthorhombic axis of lunar olivine is more compressible than that of terrestrial olivine.
\item A simple model was developed to disentangle the contributions of porosity and chemical-bond scale (e.g. lattice disorder) effects to the observed softening.
\end{highlights}



\maketitle

\section{Introduction} \label{sec:intro}
The investigation of the mechanical properties of solids is essential to develop new materials with new functionalities or to optimize the performance of existing applications. From a more fundamental perspective, understanding the mechanical and elastic properties of solids allows one to probe their atomic and molecular structure and also the relationship between microstructure (grain size, porosity, fracturing, or defects) and macroscopic behavior (performance, durability, or structural integrity). 

In particular, the study of the mechanical properties of planetary materials is currently attracting considerable research interest. In this regard, there are ongoing research lines aiming to characterize the physico-chemical properties and the mechanical response of construction materials fabricated with lunar regolith simulants for the purpose of the so-called in-situ resource utilization \citep[ISRU;][]{PILEHVAR2020,NEVES2020}. Space agencies envisage the extraction of raw materials and minerals from planetary bodies, either asteroids, moons or planets, in order to build the infrastructures required by future dwellers \citep{LIM2017_ISRU, NASER2019_ISRU, Duffard21_ISRU, LIU2024}. The first investigations on the behavior of concrete made with lunar material date back to the 80s of the last century, which relied on samples returned by the Apollo 16 mission \citep{Lin88_concreteApollo16}. 

Additionally, many of the current designed space missions involve contact processes that play a crucial role for their success. Examples of this kind are the recent landing and sample extraction on asteroids (101955)~Bennu \citep[OSIRIS-Rex mission;][]{OSIRISRex_Lauretta2017}, (25143)~Itokawa \citep[Hayabusa mission;][]{Yano06_Hayabusa}, or (162173)~Ryugu \citep[Hayabusa2 mission;][]{Watanabe19_Hayabusa2}. Contact processes are also critical for the success of lunar missions like Chang'e 7 \citep{Wang2023_Change7}, Chandrayaan-3 \citep{Chandrayaan3_DrugaPrasad2023} or the Artemis mission \citep{NASA2020_Artemis}. For instance, as shown by \citet{Tradivel14} for the case of contact motion on asteroidal surfaces, the contact dynamics strongly depends on the coefficient of restitution which, in turn, is determined by the elastic properties of the two involved bodies. 

Similarly, in the field of planetary defense, the amount of momentum transferred to a hazardous asteroid in a deflection mission is expected to be strongly dependent upon the mechanical properties of its surface, as observed in the recent DART mission \citep{Cheng2023_DART, Daly2023, Roth_2023, Thomas2023, Raducan2024}. In this sense, fragmentation modelling also requires proper knowledge of the micromechanical properties of the impacted object \citep{Davison2016, Robin2024}. 

Among the various methodologies and analysis techniques available, nanoindentation is emerging as a promising tool for assessing the mechanical performance of extraterrestrial materials, as it is a minimally damaging test that helps preserve invaluable samples. In nanoindentation, the sample is indented with a nanometer-sized sharp tip. Mechanical properties of the sample at mineral scale are extracted from the amount of indentation displacement as a function of the applied load. However, mechanical properties of individual minerals do not strictly correspond to macroscopic whole-rock properties. These can ultimately be assessed with appropriate modeling and careful characterization of fractures, as shown for the case of granite samples \citep{Xu23} and even in meteorites \citep{TANG2023}.

Nanoindentation has been widely used to investigate natural and industrial materials \citep{tabor2000hardness, Ma2020_review}, but also to characterize extraterrestrial soil simulants, including lunar ones \citep{GHOLAMI2022_lunarsoil}. To a lesser extent, it has also been applied to analyze natural extraterrestrial samples. The first works dealing with nanoindentation measurements showed that valuable information can be gained with this technique. \citet{TANG2023} attempted to model the mechanical properties of asteroid rocks based on the results of nanoindentation experiments on meteorites. Moreover, this technique has already been used to study a number of meteorites from different parent bodies \citep[e.g. ordinary and carbonaceous chondrites;][]{YomogidaMatsui83_nanoindent, Horii1990_VickersHardness, MoyanoCambero17_Chelyabinsk, PENAASENSIO2024_machinelearning}, returned samples from Itokawa \citep{Tanbakouei19_Itokawa} and Bennu \citep{Hoover_Bennu2024} asteroids, and also returned lunar rocks \citep{Nie23_Change5}.

In particular, nanoindentation tests were recently performed by \citet{PenaAsensio24_nanoindentations} in order to probe the mechanical properties of different minerals in a series of lunar meteorites. The findings of these authors suggested that, in some lunaites like the mingled regolith breccia JAH~838, the hardness ($H$) and the reduced elastic modulus ($E_r$) of the widely studied olivine mineral (i.e., the primary component of the Earth's upper mantle) might be slightly lower than in terrestrial olivine. These results would indicate that lunar and terrestrial materials might exhibit distinct mechanical behaviors. However, the differences observed in that work were not sufficiently conclusive due to limited statistical robustness.

In this work we have identified a lunaite, the NWA~12008 meteorite, which contains olivine grains with $H$ and $E_r$ values that are significantly lower than those measured in terrestrial olivine. The aim of the present work is to evaluate the mechanical properties of lunar olivine in this meteorite and compare the results with those measured on terrestrial olivines. Here we have focused our attention on olivine because the mechanical properties of this mineral have been widely studied in the past due to their relevance for the geodynamics of the lithosphere and the Earth's mantle \citep{Kranjc16_olivines, Kranjc20, BARAL2021, Badt23, Kumamoto24}, and also because it is one of the most common minerals in extraterrestrial samples \citep[e.g.][]{RubinMa17}. Similar investigations on other major minerals in lunar rocks would have been more intricate due to compositional or structural complexity (such as, for instance, the strong variability of stoichiometry in Fe/Ti/Mg oxides, or cationic substitutions in pyroxenes). 

Given that NWA~12008 is found to exhibit clearly reduced $H$ and $E_r$ values, we have performed additional high-pressure X-ray diffraction (HP-XRD) experiments in order to obtain an independent measurement of the elastic properties of olivine in this meteorite. Although HP-XRD was used in the past to investigate phase transitions of some meteorite minerals \citep{Chandra2013,Chandra16_PipliaKalan_HPXRD}, a comprehensive study of the pressure behavior of minerals in meteorites, and particularly in lunaites, is still lacking. This type of study is particularly complex because of the multiphase nature of the samples. The HP-XRD experiments suggest that the crystal lattice of the  NWA~12008 lunaite, and in particular its $b$ orthorhombic axis, may be more compressible than that of the terrestrial samples. Similar results are found in other ordinary chondrites, like the highly-shocked Chelyabinsk meteorite. A simple phenomenological model combining the nanoindentation and HP-XRD results is developed to disentangle the contributions from porosity and interatomic bonding in the observed mechanical softening behavior.

This article is organized as follows: in Section~\ref{sec:sample} we describe the analyzed samples and the different experimental techniques; the results are presented and discussed in Section~\ref{sec:results}; and, finally, the main conclusions of this work can be found in Section~\ref{sec: conclusions}.

\section{Sample description and methodology} \label{sec:sample}
NWA~12008 is a highly-shocked lunar mare basalt with a rich mineralogy that is the consequence of its igneous history. Pyroxene and olivine are the main minerals in this achondrite, as confirmed by laboratory XRD measurements on powdered material (not shown). These phases can be identified as a clinoenstatite mineral and Fe-rich forsterite, respectively. As reported in the Meteoritical Bulletin\footnote{\url{https://www.lpi.usra.edu/meteor/metbull.php}}, NWA~12008 also features sizable amounts of plagioclase and ilmenite, which are not detected by XRD. Other minerals such as barite or troilite have also been reported to be contained in NWA~12008 \citep{Meteoriticalbulletin_107_20}. 

\begin{figure}[t]
    \centering
    \includegraphics[width = 0.8\textwidth]{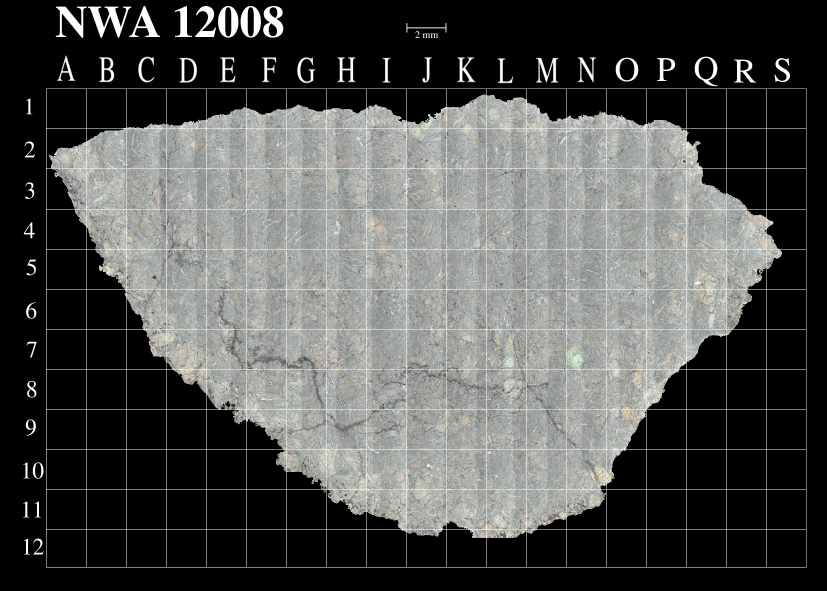}
    \caption{Mosaic of the NWA~12008 lunar meteorite sample under reflected light. Each grid separator corresponds to 2~mm in physical scale.}
    \label{fig: mosaic}
\end{figure}

A thin section of NWA~12008 was prepared, and a Zeiss Scope A1 microscope was used to build its mosaic under reflected light (Figure~\ref{fig: mosaic}). The mosaic in Figure~\ref{fig: mosaic} includes a grid with 4-mm$^2$ square regions of interest (ROIs) that are conveniently labeled for proper identification of the studied areas. The high degree of fracture of this meteorite is clearly seen in all regions of the sample (see Figure~\ref{fig: mosaic}). The most apparent feature is the large crack going across the bottom part of the image. Pyroxenes and olivines appear as submillimetric homogeneous regions, while the whitish, highly reflective minerals correspond to the opaque minerals in the sample \citep{mackenzie1994colour}. 

\subsection{Characterization using nanoindentation}\label{subsec: nanoindentation methods}
\begin{figure}[t]
    \centering
    \includegraphics[height=0.7\textheight]{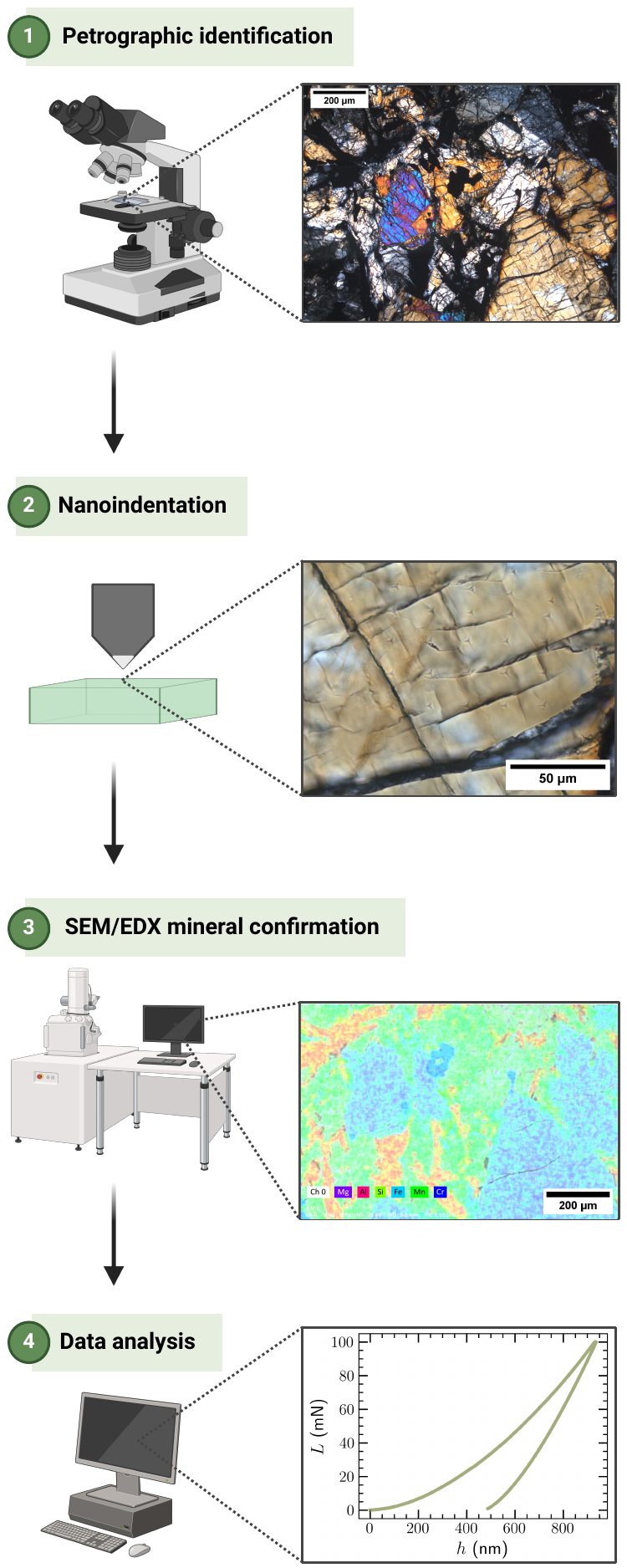}
    \caption{Methodology scheme, with images taken from NWA~12008 lunar meteorite. In this illustration, all images from Steps~1 to~3 are taken from the delimiting region between quadrants I-J~10 in Figure~\ref{fig: mosaic}. Petrographic images (Steps~1 and~2) were taken under cross-polarized light (interference colors are not standard because the thin section is not exactly \SI{30}{\micro\metre} thick). The image for Step~2 shows the NWA~12008 high olivine relief and a $5\times5$ nanoindentation matrix of the romboid-shaped mineral seen in Step~1. Elemental analysis of the same region was conducted by SEM/EDX mapping (Step~3). Each elemental abundance is represented by different color intensities: purple for \ce{Mg}, red for \ce{Al}, green for \ce{Si} and \ce{Mn}, light blue for \ce{Fe} and dark blue for \ce{Cr}. Step~4 includes an example of an applied force ($L$) vs displacement curve ($h$) from a nanoindentation in Step~3. Created in \url{https://BioRender.com}.}
    \label{fig: methodology}
\end{figure}

A summary of the analytical procedure employed in this work is shown in Figure~\ref{fig: methodology}. As nanoindentations allow retrieving the mechanical properties locally, at a submicron scale, one must carefully select the region to be nanoindented. Thus, we first conducted a visual inspection of the sample under the petrographic microscope. This allowed us to obtain a preliminary identification of the major minerals and, in particular, to identify the largest, most homogeneous olivine grains to perform the nanoindentations. For instance, the rhomboid-shaped mineral grain at the bottom right region of Figure~\ref{fig: methodology} (Step 1) was identified as an olivine grain. This grain belongs to the J10 ROI in Figure~\ref{fig: mosaic}. It should be noted that the thickness of the thin section is not the standard \SI{30}{\micro\metre} but somewhat larger ($\sim$\SI{40}{\micro\metre}), and therefore the interference colors are not the usual ones.

The selected olivine regions were nanoindented by using an Anton-Paar nanoindenter (NHT2 model). We used a diamond pyramidal-shaped with triangle-based tip, also known as Berkovich, to perform the experiments. The nanoindenter tip was calibrated using a fused silica sample \citep{FischerCripps_nanoindentation_04}. All indentation tests comprised a loading segment to \SI{100}{\milli\newton} at a \SI{200}{\milli\newton / \min}, followed by a steady pause of 2~s. Unloading was performed at the same rate. These parameters were chosen not to exceed the maximum load that the studied minerals can support, and that the maximum penetration does not exceed 1/10 the thickness of the lamella. Step 2 in Figure~\ref{fig: methodology} shows a zoom in the J10 ROI after conducting a grid of nanoindentations, where no major crystal alteration is seen after the measurements. We conducted additional nanoindentation measurements on terrestrial olivine under the same conditions, which acted as reference samples.

\subsubsection{Mineral confirmation using SEM-EDX}
In order to fully characterize the composition of the nanoindented regions and confirm that these did correspond to olivine minerals, we conducted EDX microanalyses by using a Hitachi TM4000plus tabletop SEM equipped with a Bruker EDX detector. With this technique we assured an objective mineral identification, rather than relying only on the petrographic microscope analysis. We mapped the abundance of the main elements in the field of view (Step 3 in Figure~\ref{fig: methodology}) and characterized the relative abundances in the investigated areas. The relative atomic abundances between \ce{Fe}, \ce{Mg}, and \ce{Si} allowed us to confirm that all the nanoindented regions considered in this work correspond to olivine. From the ratio between \ce{Mg} and \ce{Fe} atomic abundances (not given), we obtained the forsterite/fayalite content of the studied olivines.

\subsubsection{Nanoindentation parametrization}\label{subsubsec: nanoindentation parametrization}
A usual nanoindentation plot shows the amount of displacement of the nanoindenter ($h$) versus the applied loading force to the surface ($L$). A typical $h$-$L$ curve is represented in Figure~\ref{fig: methodology} (Step 4). Due to the material plasticity, the loading and unloading curves do not follow the same path. The whole curve can be parametrized to extract the hardness ($H$) and reduced elastic modulus ($E_r$) at the nanoindented spot, using the method described in \cite{OliverPharr_nanoindentation_2011}. 

As is well known, $H$ is the mechanical property characterizing the material resistance to permanent deformation after application of a load. It can be calculated at the point of maximum load ($L_{max}$) using the projected contact area $A_c$ on the horizontal plane as
\begin{equation}
    H = \frac{L_{max}}{A_c}.
\end{equation}
For an ideal Berkovich nanoindenter, the contact area is ${A_c = 24.5 h_c^2}$, where $h_c$ is the contact depth of the tip.

On the other hand, the Oliver-Pharr method allows calculating $E_r$, which is directly related to the contact stiffness, $S$, which, in turn, is given by the slope of the load-displacement curve (Step 4 plot in Figure~\ref{fig: methodology}) as 
\begin{equation}
S = \frac{dL}{dh} = 2\beta \sqrt{\frac{A_c}{\pi}}E_r,
\end{equation}
where $\beta = 1.034$ is a constant related to the geometry of the Berkovich indenter \citep{FischerCripps_nanoindentation_04}.

The reduced elastic modulus characterizes the elastic deformation of the material under an applied force. This parameter is directly measured by the nanoindenter and includes the contributions from the elastic moduli of both the nanoindenter tip (diamond) and the material. The reduced elastic modulus can be related to Young's modulus of the measured sample with further parametrization \citep{OliverPharr_nanoindentation_1992}.

The areas below the loading and unloading curves correspond to the total ($W_{T}$) and plastic ($W_{el}$) works during the indentation process, respectively. The plastic energy involved in the process can be found as ${W_{pl} = W_T - W_{el}}$. These values provide insights on the recovery stage after the nanoindentation, i.e. on the reversible and irreversible energies exchanged during the process \citep{FischerCripps_nanoindentation_04, Kaushalendra12, Liu16_nanoindentation_shalerocks}. In this work, we used the plastic and elastic works as complementary parameters to identify samples near to cracks or grain boundaries (Section~\ref{subsec: nanoindentation results}).

\subsection{HP-XRD} \label{sec: HP methods}
Room-temperature powder angle-dispersive XRD measurements were performed on different meteorite samples as a function of hydrostatic pressure. All samples were carefully powdered and loaded with a 4:1 methanol-ethanol mixture in a membrane-type diamond anvil cell (DAC) with diamond culets of \SI{400}{\micro\metre} in diameter. The measurements were performed during the upstroke and below 12~GPa, since the methanol-ethanol pressure transmitting medium exhibits an abrupt reduction of hydrostaticity above this pressure value. 

The HP-XRD experiments were performed in the BL04-MSPD beamline at ALBA synchrotron, Cerdanyola del Vallès, Barcelona \citep{Fauth_ALBA13}. This beamline is equipped with Kirkpatrick-Baez mirrors, which allow focusing the monochromatic beam (wavelength of 0.4246~\AA, corresponding to the absorption K-edge of Sn), and a Rayonix CCD detector with an active area of 165-mm diameter. 

The samples included in the HP-XRD study consisted of the NWA12008 meteorite together with 3 additional ordinary chondrites, which served us to evaluate the compression behavior in other non-terrestrial materials: Aiquile (classified as H5), Chelyabinsk (LL5), and Vi\~{n}ales (L6). The pressure applied on the samples was determined with the equation of state (EoS) of copper \citep{Dewaele_EoS04}. The associated error to the pressure determination, including the pressure gradients inside the DAC, is lower than 0.2~GPa. The sample-to-detector distance, along with various detector geometrical parameters of the experiment, were calibrated with the DIOPTAS software \citep{Prescher15_DIOPTAS}. For this purpose, diffraction data from \ce{LaB6} was used. 

In order to obtain enough XRD signal from reflections of the mineral olivine in the studied samples, many different XRD measurements were performed at each applied pressure. This was achieved by mapping the sample cavity. For the subsequent analysis of the data, only the best scans were employed. This was particularly important in the case of NWA~12008, since the overall olivine content of this meteorite is much lower than that of the ordinary chondrites.

Structural analyses for the olivine mineral contained in the studied samples were performed with the program TOPAS~4.2 from Bruker. Most of the analyses relied on a full-pattern matching based on the Pawley method, which allowed us to obtain the unit-cell parameters of olivine as a function of hydrostatic pressure. EoS were fitted to the pressure-volume or pressure-lattice parameter data by using EosFit-GUI \citep{Gonzalez-Platas:kc5039}, which allowed us to extract the compressibility properties of olivine in the investigated samples. 

\section{Results and discussion} \label{sec:results}

Olivine, \ce{(Mg,Fe)2SiO4}, is one of the most common minerals in the Solar System. It is a solid solution between forsterite (Fo), \ce{Mg2SiO4}, and fayalite (Fa), \ce{Fe2SiO4}. Olivine is one of the main minerals in the Earth's mantle and has been widely studied in the past due to its relevance in geodynamics \citep[e.g.][]{Mackwell90, Peslier10}. In other planetary bodies, such as Venus or Mars, olivine has been found to be a major component of their mantles \citep{MorganAnders80, Fegley03, Taylor03, Ronco15}. Smaller bodies such as chondrite meteorites also contain olivine \citep[][and references therein]{RubinMa17}. In fact, olivine is the main mineral in interstellar dust particles, formed during the last stages of stellar evolution \citep{Jones00, Min07}. In the present work, we use nanoindentations and HP-XRD measurements to investigate the mechanical and elastic properties of olivine in samples from different origins. 

\subsection{Nanoindentation results} \label{subsec: nanoindentation results}

A total of 78 nanoindentations were performed on lunar olivine grains of 5 different ROIs of the NWA~12008 thin section. Due to the high degree of fracture of the sample, some nanoindentations were conducted too close to visible microfractures or grain boundaries, which are not representative of the whole mineral \citep{Avadanii23}. In particular, three of the measurements were found to clearly deviate from the general behavior (including $H$, $E_r$, $W_{el}$, and $W_{pl}$) and were discarded.

In turn, 91 nanoindentation measurements were performed on terrestrial olivine. The selected sample showed grains with uniform chemical composition and no micrometric fractures, contrary to the case of NWA~12008. Therefore, no outlier removal was required.

\begin{figure}[t]
    \centering
    \includegraphics[width = 0.75\textwidth]{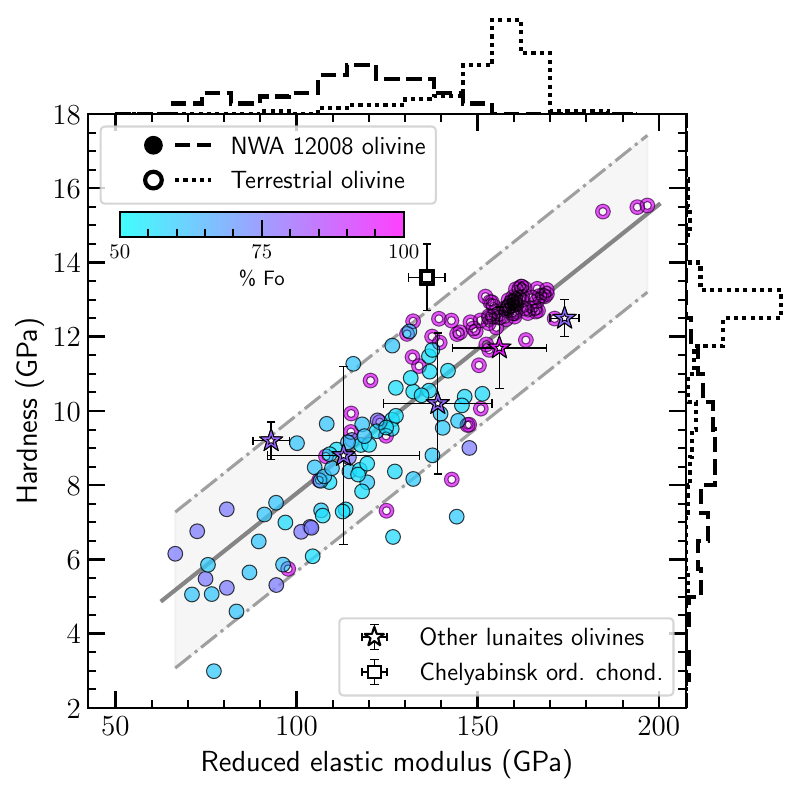}
    \caption{Values of hardness and reduced elastic modulus for each individual nanoindentation in this work. Closed circles correspond to NWA~12008 lunar olivine data, while open circles correspond to terrestrial olivine values. The solid line corresponds to the all-data linear fit. The region delimited with dash-dotted lines represents the 95\% prediction interval of the data. The marginal histograms show the density distribution of terrestrial and NWA~12008 values along the hardness and reduced elastic modulus spaces. Values from olivines in DHOFAR~1084, JAH~838 and NWA~11444 lunar meteorites \citep[stars,][]{PenaAsensio24_nanoindentations} are also shown, as well as those from the Chelyabinsk meteorite \citep[square,][]{MoyanoCambero17_Chelyabinsk}, along with their reported errors. Each dot is colored according to the percentage of forsterite (\%Fo) in the olivine except for the Chelyabinsk meteorite, for which no \%Fo was reported.}
    \label{fig: H vs E}
\end{figure}

Figure \ref{fig: H vs E} shows the measured values resulting from the individual nanoindentations on NWA~12008 and terrestrial olivines in a $H$~vs.~$E_r$ plot. The histograms on the top and right sides of the figure show the distribution of $H$ and $E_r$ values obtained for NWA12008 (dashed lines) and for terrestrial olivine (dotted lines). As can be seen in the figure, most NWA~12008 olivine data points (solid dots) exhibit significantly lower hardness and reduced elastic modulus compared to their terrestrial counterparts (open dots), which cluster around $H\sim 13$~GPa and $E_r\sim 160$~GPa. These values are in good agreement with those reported in the literature for terrestrial olivine \citep{Kranjc16_olivines}. Table \ref{tab: nanoindentation results} shows a compilation with the mean values of $H$ and $E_r$ for the lunar (NWA~12008) and terrestrial olivines measured in this work. The errors reported in the table represent the standard deviation of the mean values. Clearly, the data points from NWA~12008 tend to populate an area with considerably lower $H$ and $E_r$ values, thus suggesting that the lunar olivine in this meteorite is softer and more elastic.

\begin{table}[b]
\centering
\caption{Mean values for hardness ($H$) and reduced elastic modulus ($E_r$) for NWA~12008 and terrestrial olivines. Experiments were conducted with a Berkovich tip with at \SI{200}{\milli\newton / \min} load rate, reaching a maximum load of \SI{100}{\milli\newton} followed a steady pause of \SI{2}{\second}. Uncertainties correspond to 1$\sigma$ standard deviation of the measurements.}
\label{tab: nanoindentation results}
\begin{tabular}{lcc}
\toprule
 & \textbf{$\boldsymbol{H}$ (GPa)} & \textbf{$\boldsymbol{E_r}$ (GPa)} \\ \midrule
\multicolumn{1}{c}{\textbf{NWA~12008 olivine}} & \textbf{$8.4 \pm 1.9$} & \textbf{$114 \pm 21$} \\
\textbf{Terrestrial olivine} & \textbf{$12.2 \pm 1.6$} & \textbf{$153 \pm 17$} \\
\bottomrule
\end{tabular}%
\end{table}

Figure \ref{fig: H vs E} also shows $H$ and $E_r$ values in olivine grains of other extraterrestrial sources, as reported in \citet{MoyanoCambero17_Chelyabinsk, PenaAsensio24_nanoindentations}. The plot also includes the 95\% prediction interval for all the measured data (lunar and terrestrial), which displays a linear trend with a $H/E_r$ ratio equal to $0.0778 \pm 0.0030$. This value is only slightly lower than that reported in \citet{Kranjc16_olivines} for terrestrial olivine ($0.0790\pm 0.0040$). The hardness of the Chelyabinsk olivine is of the order of the terrestrial values, while its reduced elastic modulus is slightly lower than that from terrestrial olivine. On the other hand, all the $H$ and $E_r$ data for the lunaites studied in \citet{PenaAsensio24_nanoindentations} lie within the 95\% prediction interval of the linear regression. Although the measurements of \citet{PenaAsensio24_nanoindentations} employed slightly different nanoindentation parameters, the data points from the lunaites studied in that work match well with the present data on NWA~12008. Their work did not provide concluding evidence for softening, as found here for NWA~12008, due to limited statistics.    

The distribution of $H$ and $E_r$ values in NWA~12008 olivine shown in Figure~\ref{fig: H vs E} is significantly more scattered compared to that measured in the terrestrial sample. We attribute this observation to relatively large differences in the particular environment of the minerals within the meteorite slab \citep{Angel2014, Angel2015, MAZZUCCHELLI2019}. Nevertheless, the overall $H$ and $E_r$ values measured for the olivine grains in NWA~12008 and also in other lunaites (see \citet{PenaAsensio24_nanoindentations} for details) are clearly lower than those in the terrestrial sample. 

It is important to remark that the fact that lunar olivines have lower $H$ and $E_r$ than their terrestrial counterparts does not seem to be a consequence, at least solely, of the particular Mg/Fe (Fo/Fa) composition of the nanoindented olivine grains. The color scale in Figure~\ref{fig: H vs E} serves to represent the molar content of Fo in the analyzed olivines. The measured lunar olivines cover a Fo range between 55\% and 75\%, which is lower than the forsterite content in the analyzed terrestrial sample ($\text{Fo}=89$\%). At first glance, one might conclude that this might explain why terrestrial olivine is harder and less elastic than lunar olivine. However, it is worth mentioning that the mechanical properties of the lunaite olivines do not seem to exhibit any monotonic behavior as a function of Fo. In fact, olivines in the lunar sample seem to even show decreasing values of $H$ and $E_r$ with increasing forsterite content. For instance, numerous olivine grains of NWA~12008 with $\text{Fo}\sim 75$\% show much lower $H$ and $E_r$ values than other grains with higher Mg contents. The fact that many data points from lunar olivine populate a region below $H=8$~GPa and $E_r=100$~GPa cannot be solely explained by a compositional effect.  

To further support this conclusion, we note that the Vickers hardness ($H_V$) and the shear stress modulus ($G$) of polycrystalline materials exhibit a universal trend, as discussed in \citet{Chen11_VickersHardness}. For the present work, and given the similarities and comparable $H$ values extracted with Vickers and Berkovich experiments \citep{Chudoba06}, it can be assumed that the data of \citet{Chen11_VickersHardness} can be extrapolated to the present case. As shown in that work, $H_V$ and $G$ in intrinsically-brittle materials can be assumed to be linear, with $H_V = 0.151G$. Using this expression together with data from \citet{bass1995elasticity}, where the shear stress of pure forsterite and fayalite in terrestrial olivine was characterized, it is possible to evaluate the expected $H$ values for olivines with $\text{Fo}\sim50\%$, i.e., for the lowest Mg contents measured in NWA~12008 olivines. These authors found that pure forsterite has ${G=81.1}$~GPa, while ${G=50.7}$~GPa in pure fayalite. As a rough approximation, we can consider that the shear stress for a 50\%Fo olivine is $\sim 66$~GPa, in agreement with previous results reported in \citet{Chung1971} ($G=67.4$~GPa). According to the relation between $H_V$ and $G$ given in \citet{Chen11_VickersHardness}, using the $H$ values that we measure in terrestrial olivine, $G=66$~GPa would correspond to a hardness of 9.9~GPa for $\text{Fo}=50\%$. This value is still much larger than $H$ in NWA~12008 olivine, where much lower values are most frequently found (Figure~\ref{fig: H vs E}). The present estimation of $H$ based on the data of \citet{Chen11_VickersHardness} supports the idea that the differences between lunar and terrestrial olivines cannot be solely attributed to compositional effects.

\begin{figure}[t]
    \centering
    \includegraphics[height=0.7\textheight]{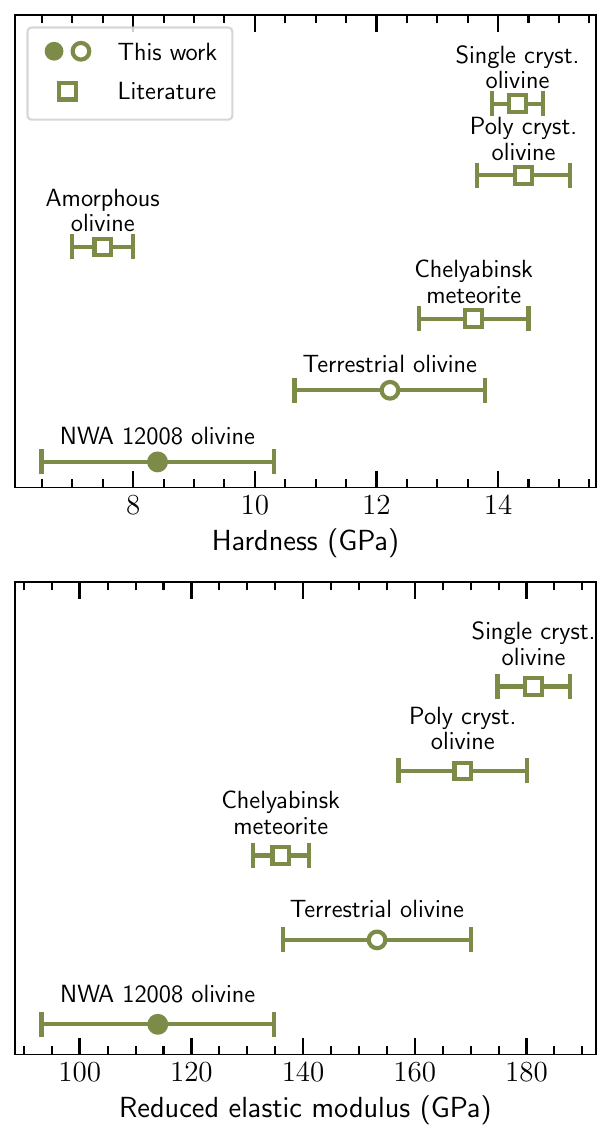}
    \caption{Mean measured values of olivine hardness and reduced elastic modulus in different sources. Markers distinguish literature values from those values extracted from nanoindentations of this work. As in Figure~\ref{fig: H vs E}, closed circles represent the NWA~12008 lunar olivine values, while open circles correspond to terrestrial olivine mean values. Single crystal and polycristal olivine values are extracted from \cite{Kranjc16_olivines}, while the amorphous olivine values are taken from \cite{BARAL2021}. Chelyabinsk meteorite nanoindentation values are from \cite{MoyanoCambero17_Chelyabinsk}. Error bars correspond to 1$\sigma$ standard deviation.}
    \label{fig: H and E}
\end{figure}

Figure \ref{fig: H and E} shows average $H$ and $E_r$ values obtained in this work for lunar and terrestrial olivines. For comparison, values from different types of terrestrial olivines (single crystal, polycrystalline and amorphous) reported in the literature \citep{Kranjc16_olivines, BARAL2021} are also plotted in the figure, together with data for the Chelyabinsk meteorite \citep{MoyanoCambero17_Chelyabinsk}. The error bars indicate the standard deviation for each collection of measurements. Note that different nanoindentation parameters were used in different works. In any case, as can be clearly seen in Figure~\ref{fig: H and E}, lunar olivines in NWA~12008 are significantly softer and more elastic than the rest of (poly)crystalline olivines \citep{Kranjc16_olivines}, including those in the highly-shocked Chelyabinsk chondrite. Error bars for NWA~12008 do not overlap with the rest of crystalline olivine samples. Interestingly, amorphous olivine \citep{BARAL2021} was found to exhibit much lower hardness values, comparable to those measured in NWA~12008. 

Up to this point, the reason for the softening of NWA~12008 and other lunaites is still an open question and could be attributed to many different phenomena, ranging from lattice disorder to the presence of (sub)micrometric features like microfractures or increased porosity. Any of these might have been caused by alterations due to shock compaction and fracturing produced by collisional processing, a key process in space weathering \citep{Bennett2013}. Moreover, \citet{JIANG2024} found that $H$ and $E_r$ in olivine are reduced with increasing density of atomic hydrogen in the crystal lattice, simulating a solar wind effect. This effect might be more pronounced with large exposure times and even due to geological aging. With regard to this, however, recent work on lunar glasses has found that geological time scales’ aging can lead to unusually huge modulus enhancements \citep{Chen23_younglunar}, contrary to the present observations. The fact that amorphous olivine is much softer than the (poly)crystalline mineral suggests that lattice disorder might have a bearing on the mechanical properties of olivine. Such disorder might include defects in various dimensions including point, line and planar defects and even 3D inclusions \citep{Demouchy21_defectsolivine}. The latter would have been introduced in the original lunar magmas.

With regard to the possible effect of microfractures on the mechanical behavior of olivine, it is interesting to compare the $H$ and $E_r$ values for olivine in the lunaites with those measured in the Chelyabinsk meteorite (Figure~\ref{fig: H and E}). This ordinary chondrite is highly shocked and contains a large number of microfractures and even darkened areas \citep{MoyanoCambero17_Chelyabinsk}. However, the measured $H$ and $E_r$ values in NWA~12008 are still much lower than those in Chelyabinsk, the mechanical behavior of which is not far from terrestrial (poly)crystalline olivine. This result seems to indicate that shock-induced microfractures are not the main responsible for the softening observed in NWA~12008. 

In relation to porosity, it is well known that, in many different types of materials, it can strongly affect $H$ and, especially, $E_r$ \citep[e.g.][]{RamakrishnanArunachalam1993, BrantleyMellott00_silicateporosity, Biener04, Pellicer12, Tolu13, Tanbakouei19_Itokawa,Cariou2008, Magoriaec09_porositySi}. Indeed, porosity implies morphological variations that may modify the mechanical behavior of materials at the submicron scale (i.e., the scale involved in the nanoindentation experiments), which is a consequence of the fact that void spaces do not contribute to the material’s resistance to loads. This yields overall softening as well as reduced sound velocities. Thus, it can be hypothesized that some extraterrestrial olivines may exhibit reduced $H$ and $E_r$ values due to increased porosity. With regard to this, \citet{Macke10, Macke11} measured the porosity of a large number of chondrites and achondrites. For the particular case of lunaites (4 samples included in their study), these authors found highly scattered porosity values, but in most cases comparable to those of ordinary chondrites \citep{Macke11}. Unfortunately, no porosity data are available for NWA~12008. In any case, it would be desirable to perform nanoindentations in different meteorites with known porosities, as this would allow one to correlate softenings and porosities. In particular, highly-porous achondrites like howardites \citep{Macke11} would be very useful in order to ascertain the effect of porosity on the nanoindentation behavior of extraterrestrial olivines. The possible effects of porosity and atomic bonding on mechanical properties are discussed in more detail in Section~\ref{subsec: model}.

\subsection{HP-XRD measurements} \label{subsec: HP-XRD results}

In the previous section, we have found strong evidence for reduced values of $H$ and $E_r$ in olivine grains of NWA~12008. Nanoindentation experiments have allowed us to obtain information about the elastic properties of lunar and terrestrial olivines, particularly $E_r$, by measuring the stiffness of the contact between the tip and the studied material. It must be stressed that the nanoindentation measurements are determined by both the macroscopic and the chemical-bond scales in a quasi-static regime.

In turn, HP-XRD measurements allow one to probe the compressibility of crystalline materials at the chemical bond scale through the determination of the unit-cell parameters as a function of pressure. By fitting the resulting pressure-volume EoS, parameters like the isothermal bulk modulus can be obtained. The bulk modulus, defined as ${B_0=-VdP/dV}$, where $P$ is pressure and $V$ is the unit-cell volume, can be understood as the resistance of a solid material to bulk compression. Bulk modulus can be related to other elastic properties such as Young's modulus. 

In the case of porous materials, the macroscopic porosity, understood here as open-shell cavities and voids that do not affect the strain distributions on the material, is not expected to strongly affect the compression behavior observed in the HP-XRD measurements. Therefore, HP-XRD may be useful to probe the intrinsic elastic properties of the extraterrestrial olivines at structural level and disentangle the role of porosity on the nanoindentation results.

In this section, we present synchrotron-based HP-XRD measurements to determine $B_0$ of olivine in NWA~12008 and compare the resulting value with that of terrestrial olivines. For comparison purposes, we have also performed HP-XRD measurements on three different ordinary chondrites.

\begin{figure}[t]
    \centering
    \includegraphics[height=0.7\textheight]{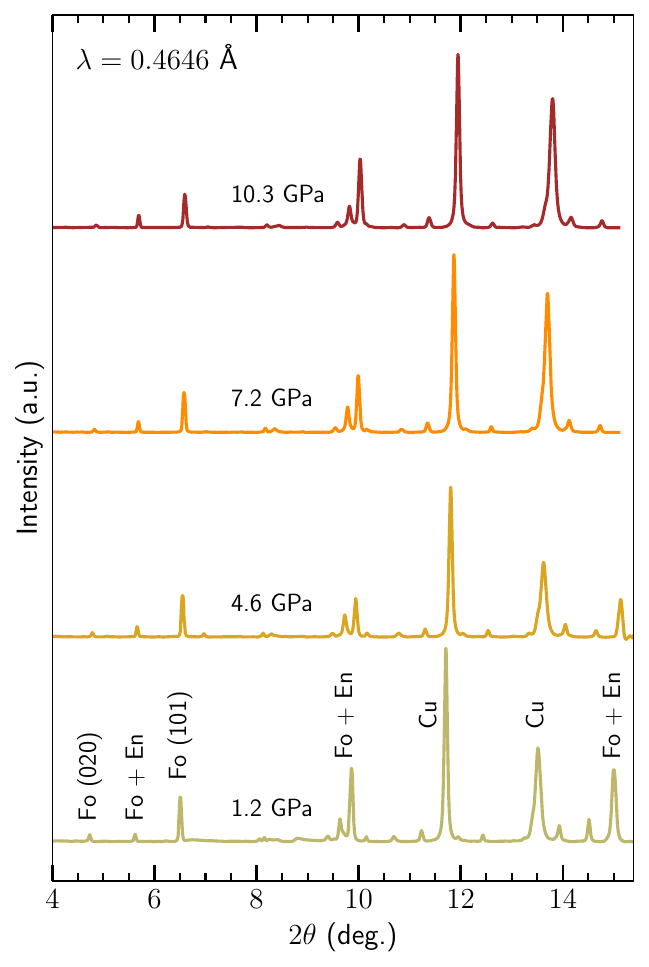}
    \caption{HP-XRD scans of lunaite NWA~122008 at different pressures. Selected reflections of forsterite (Fo) and clinoenstatite (En) are labelled in the lower scan. Peaks from copper are also indicated.}
    \label{fig: HP-XRD}
\end{figure}

Figure ~\ref{fig: HP-XRD} shows selected powder XRD scans obtained for NWA~12008 at different pressure values. The expected shift of all the observed reflections to high $2\theta$ values (i.e. to low lattice spacings) with increasing pressure is clearly seen. From the point of view of HP-XRD experiments, these are standard results and, qualitatively, they are very similar to those widely reported in the literature for terrestrial olivine by means of single-crystal XRD (sc-XRD) and powder XRD \citep[][and references therein]{Nestola11_EPS,Nestola11}. In the case of the rest of ordinary chondrites included in the present study (scans not shown for simplicity), similar results were obtained. In fact, the particular compression behavior of different types of samples (terrestrial, lunar, and asteroidal) with similar mineralogy only emerges after a careful analysis of the experimental data.

\begin{figure}[t]
    \centering
    \includegraphics[height=0.7\textheight]{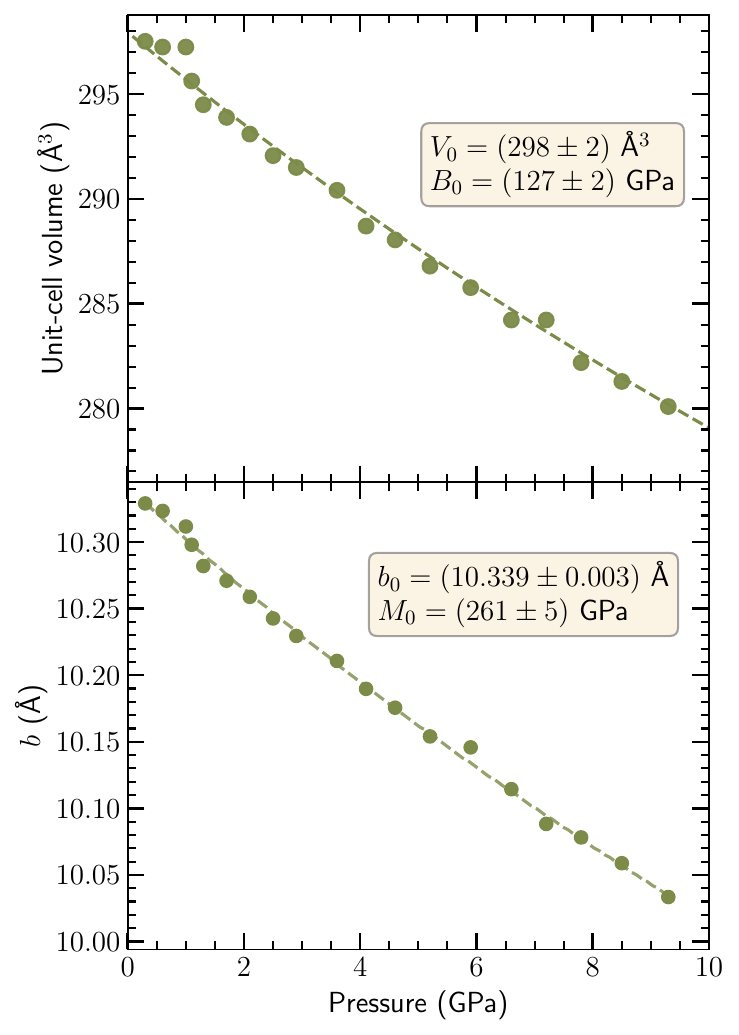}
    \caption{Upper panel: pressure-volume dependence (solid dots) as obtained from HP-XRD measurements for lunaite NWA~12008. The curve shows the result of a fit to the experimental data with a third-order Birch-Murnaghan equation of state. Lower panel: pressure dependence of the $b$ lattice parameter of olivine (solid dots) as extracted from the evolution of the (020) reflection. The curve shows the result of a fit to the experimental data with a second-order Birch-Murnaghan equation of state.}
    \label{fig: EoS}
\end{figure}

The Pawley method was used to extract the orthorhombic unit-cell parameters of olivine as a function of pressure for the different meteorites measured by HP-XRD. The upper panel of Figure ~\ref{fig: EoS} shows the resulting unit-cell volume, as a function of pressure, for the particular case of NWA~12008. Data up to 9.5 GPa is shown in the figure, together with a fit using a third-order isothermal Birch-Murnaghan EoS (BM-3):

\begin{equation}
    P(V) = \frac{3B_0}{2} \Bigg[ \left( \frac{V_0}{V} \right)^{7/3} - \left( \frac{V_0}{V} \right)^{5/3}\Bigg] \cdot \Bigg\{1+\frac{3}{4}(B'-4)\left[\left(\frac{V_0}{V}\right)^{2/3}-1\right]\Bigg\}, 
    \label{eq: BirchMurnaghan EoS}
\end{equation}
where $V_0$, $B_0$ and $B'$ are, respectively, the zero-pressure unit-cell volume, the zero-pressure bulk modulus and its pressure derivative. Normalized pressure-eulerian strain plots (not shown) indicated that $B'$ is close to 4, thus suggesting that a second-order BM EoS (BM-2, where $B'=4$ in Equation~\ref{eq: BirchMurnaghan EoS}) might suffice for the EoS fits. However, as previous works that obtained accurate EoS parameters for olivines with sc-XRD measurements relied on BM-3 \citep{Nestola11_EPS, Nestola11}, here we fit our experimental HP-XRD data to a BM-3 EoS (Equation \ref{eq: BirchMurnaghan EoS}). For this purpose, the well-known correlation between the fitted $B_0$ and $B'$ parameters for BM-3 EoS \citep{Angel00} needs to be considered. This correlation typically leads to large uncertainties in the fitted values. To avoid this problem, we fixed $B'$ to the expression ${B'=31.104-0.0874V_0}$, which was obtained by adjusting the $B'$--$V_0$ dependence reported in \citet{Nestola11} to a linear equation. As shown in that work, $B_0$ and $V_0$ in terrestrial olivines increases with Fe content, while $B'$ tends to slightly decrease.

\begin{table}[h]
    \centering
    \caption{Summary of the experimental results obtained with HP-XRD. For each analysed meteorite (NWA~12008 is lunaite and the rest are ordinary chondrites), we estimated the bulk modulus of olivine ($B_0$) together with its unit-cell zero-pressure volume ($V_0$) and its linear axial bulk modulus for the $b$ orthorhombic lattice parameter ($M_b$), together with the zero-pressure $b$ lattice-parameter ($b_0$). $B_0$ and $V_0$ were extracted from fits to a BM-3 EoS by fixing B' to data in the literature, while $M_b$ and $b$ were obtained from fits to a BM-2 EoS (see main text).}
    \label{tab: HP-XRD results}
    \begin{tabular}{lcccc}
        \toprule
         \textbf{Meteorite} & $\boldsymbol{B_0}$ \textbf{(GPa)} & $\boldsymbol{V_0}$ \textbf{(\AA$^3$)}& $\boldsymbol{M_b}$ \textbf{(GPa)} & $\boldsymbol{b_0}$ \textbf{(\AA)} \\ \midrule
         NWA~12008 & $127\pm 3$ & $298.0 \pm 0.2$ & $261\pm5$ & $10.339\pm0.003$ \\
         Chelyabinsk & $122 \pm 1$ & $296.0 \pm 0.1$ & $270 \pm 6$ & $10.285 \pm 0.003$ \\
         Aiquile & $124 \pm 2$ & $294.1 \pm 0.2$ & $281 \pm 8$ & $10.260 \pm 0.015$ \\
         Vi\~{n}ales & $127 \pm 1$ & $294.6\pm0.1$ & $273 \pm 5$ & $10.268 \pm 0.004$  \\ \bottomrule
    \end{tabular}
\end{table}

\begin{figure}[t]
    \centering
    \includegraphics[width = 0.75\textwidth]{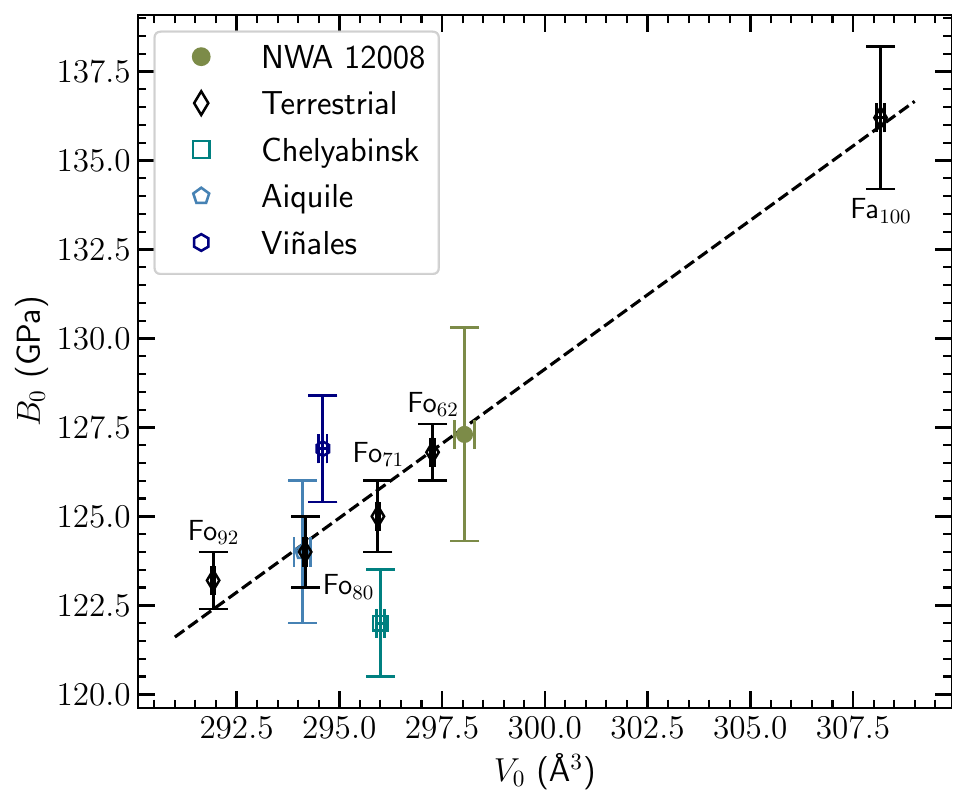}
    \caption{Bulk modulus ($B_0$) versus zero-pressure volume ($V_0$) for olivine in the different samples measured in this work by HP-XRD. The $B_0$ and $V_0$ values were obtained by fitting the experimental pressure-volume data of Figure \ref{fig: EoS} (upper panel) with a third-order Birch Murnaghan EoS. Data for terrestrial olivines is taken from \citet{Nestola11_EPS, Nestola11} and \citet{Zhang1998} and their respective \%Fo is labeled accordingly. Terrestrial olivine data was interpolated from a linear fit, relating $B_0 = 0.84 V_0 - 121.7$ GPa.}
    \label{fig: B_0}
\end{figure}

Figure \ref{fig: B_0} shows a compilation of the $V_0$ and $B_0$ values extracted with BM-3 fits (fixing $B'$ as explained above) for the different meteorite samples included in the present HP-XRD study. The corresponding data is summarized in Table~\ref{tab: HP-XRD results}. For comparison purposes, the $B_0$ values reported in \citet{Nestola11_EPS,Nestola11} are also shown in Figure \ref{fig: B_0}. A sc-XRD data point for a fayalite sample \citep{Zhang1998} has also been included in the plot. A linear regression was performed to these reference data, relating $V_0$ and $B_0$ as $B_0 = 0.84 V_0 - 121.7$ GPa. It is important to remark that all these sc-XRD data for the (terrestrial) forsterite-fayalite solid solution were obtained below $\sim 10$ GPa, i.e., in a pressure range comparable to that of the present powder HP-XRD measurements. 

Among the ordinary chondrites, only the Chelyabinsk meteorite seems to exhibit increased compressibility (i.e., reduced $B_0$) relative to the terrestrial samples, which is consistent with the slightly reduced $E_r$ values measured in this highly shocked meteorite by nanoindentation (see Figure~\ref{fig: H and E} and \citet{MoyanoCambero17_Chelyabinsk}). In contrast, the bulk modulus obtained for olivine in NWA~12008 seems to be comparable, within experimental error, to that measured in terrestrial olivines of similar Mg content (Figure \ref{fig: B_0}). 

However, the previous analysis of the XRD scans with the Pawley method is strongly limited by the multiphase nature of the meteorite samples, as shown in Figure~\ref{fig: HP-XRD} for NWA~12008, in which both enstatite and forsterite peaks have been detected. In the particular case of this sample, the total olivine content is relatively low, and most of the reflections from olivine overlap to peaks from the pyroxene phase, the major mineral in this meteorite. As can be seen in Figure~\ref{fig: HP-XRD}, only the (020) and (101) reflections of olivine are not interfered by peaks from the pyroxene phase. Unfortunately, these two peaks alone do not allow extracting the pressure behavior of the three orthorhombic lattice parameters of olivine ($a$, $b$, and $c$).

To overcome this problem, we have performed an alternative analysis of the powder HP-XRD scans in which we have solely focused on the evolution of the (020) reflection of olivine, as this peak does not overlap with any other reflections and provides direct information about the compression behavior of the $b$ parameter. Interestingly, this lattice parameter corresponds to the most compressible axis of olivine \citep{Nestola11} and, therefore, it may be more sensitive to structural changes of the crystal lattice of this mineral.  

The bottom panel of Figure~\ref{fig: EoS} shows the pressure behavior of $b$ of olivine in NWA~12008 as determined from the evolution upon compression of the (020) reflection of this mineral. For simplicity, we extracted the zero-pressure axial bulk modulus of $b$, $M_b\equiv3B_b=-bdP/db$, which accounts for the resistance to compression along the $b$ axis, by fitting the experimental data to a second-order BM-2 (i.e., fixing ${M_b' = 12}$). For this sample, the fitting returns a value ${M_b=(261 \pm 5)}$~GPa, which is much lower than that reported in \citet{Nestola11} for all the olivine samples studied in that work, up to $\text{Fa}=38\%$ ($M_b=279$--282~GPa).

\begin{figure}[t]
    \centering
    \includegraphics[width = 0.75\textwidth]{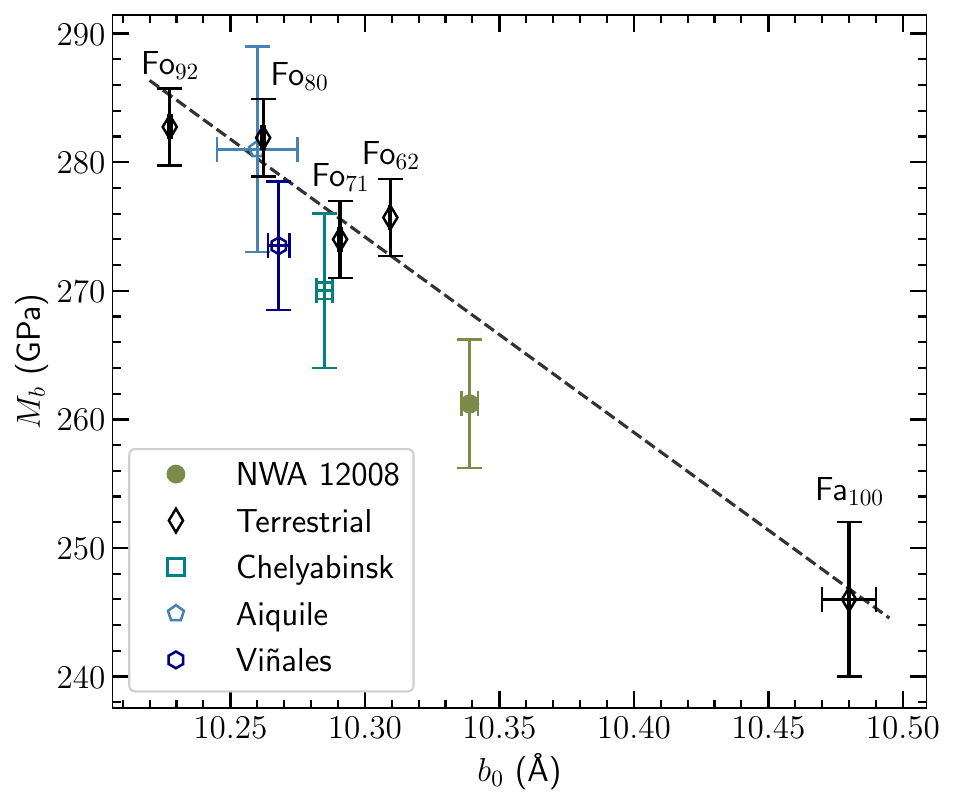}
    \caption{Linear axial bulk modulus for the orthorhombic $b$ lattice parameter of olivine ($M_b$) versus the zero-pressure lattice-parameter ($b_0$) for the different samples measured in this work by HP-XRD. The $M_b$ and $b_0$ values were obtained by fitting the experimental pressure-lattice parameter data with a second-order Birch-Murnaghan EoS. Data for terrestrial olivines is taken from \citet{Nestola11} and \citet{Zhang1998}. A linear fit to the data from the terrestrial olivines gives $M_b = -152.1 b_0 + 1840.4$ GPa.}
    \label{fig: M_0}
\end{figure}

Note that the data of \citet{Nestola11_EPS,Nestola11} were obtained with a BM-3 EoS. In this case, no compositional trend was found for the pressure derivative of $M_b$. In order to properly compare between the $M_b\equiv3B_b$ values reported in that work and those of the present study, we have repeated the fits to the data of \citet{Nestola11_EPS, Nestola11} by using a BM-2 EoS. Figure \ref{fig: M_0} shows the resulting EoS parameters for the olivine samples of that work, together with the results that we find for our meteorite samples, which are also summarized in Table~\ref{tab: HP-XRD results}. A data point for fayalite from \citet{Zhang1998}, also fitted with a BM-2 EoS, is included in the plot. As in Figure~\ref{fig: B_0}, we include a linear fit of the terrestrial data describing the dependence of $M_b$ on $b_0$ as ${M_b = -152.1 b_0 + 1840.4}$~GPa.

As can be seen in Figure~\ref{fig: M_0}, the $M_b$ values for the $b$ axis in the terrestrial olivines decrease with increasing fayalite content. This is a well-known result in the case of terrestrial, high-quality olivines: while the $a$ and $c$ parameters of olivine become less compressible with increasing Fa, the $b$ parameter becomes more compressible \citep{Zhang1998, Nestola11}. Overall, the forsterite-fayalite solid solution becomes less compressible with increasing Fe content, as shown in Figure~\ref{fig: B_0}.

For the meteorite samples studied in this work, the $b$ axis of olivine in several of the samples seems to be more compressible than expected for their particular composition, within experimental error. In the case of NWA~12008, there seems to be a significant reduction of the $M_b$ value, even after comparing with the \citet{Nestola11} data refitted to a BM-2 EoS (note that much larger differences would be seen without the BM-2 refitting). For the Chelyabinsk meteorite, the resulting $M_b$ value also seems to be lower, in agreement with the behavior already observed for $B_0$ (Figure~\ref{fig: B_0}). In the case of Vi\~{n}ales, a reduced $M_b$ value is also found. For the Aiquile meteorite, the obtained $M_b$ is in agreement with that from terrestrial olivine reference data. No nanoindentations are available for these last two meteorites, for which further discussion on their mechanical behavior is not possible. 

\subsection{Porosity vs. chemical-bond softening}\label{subsec: model}

All these results for the axial compressibility of the $b$ parameter of olivine in NWA~12008 and Chelyabinsk, in combination with the nanoindentation data available for these two meteorites, indicate that the elastic properties of certain planetary materials may be influenced by the various processes they underwent throughout their complex histories. For example, these meteorites could have increased porosities, which could be expected to yield elastic/mechanical softening \citep{BARLA1984, RamakrishnanArunachalam1990, RamakrishnanArunachalam1993}. Also, shock metamorphism \citep{TRIGORODRIGUEZ09_shocks, Beitz16_shocks} or even cosmic-ray bombardment \citep{Poppe_2023, JIANG2024} could have affected the crystalline quality of some of the mineral constituents, thus altering the mechanical behavior at the crystal lattice scale. Regardless of its origin, the observed softening may be particularly relevant from an applied point of view, namely for ISRU applications, as the material response of lunar or asteroidal rocks might be sizably different from that of their terrestrial counterparts, making current simulant materials non-reliable to test processes and materials for future lunar surface engineering construction.

In order to shed additional light on the effect of increased (macroscopic) porosity on the mechanical properties of our samples, we have used the simple model of \citet{RamakrishnanArunachalam1990}, which allows one to calculate the Young's modulus of a porous solid with randomly-distributed macroscopic pores ($E_{\text{por}}$) relative to the non-porous (bulk) reference material ($E_\text{ref}$) as:
\begin{equation}
\frac{E_{\text{por}}}{E_\text{ref}}=\frac{(1-\varphi)^2}{1+\kappa_E\varphi},
\label{eq: por1}
\end{equation}
where $\varphi$ is the porosity fraction of the porous material. In the previous equation, the parameter $\kappa_E$ depends on the Poisson's ratio ($\nu$) for the non-porous solid as
\begin{equation}
\kappa_E=2-3\nu.
\label{eq: por2}
\end{equation}
Similar equations can be written for the bulk modulus and the shear modulus of the porous solid as a function of $\varphi$ \citep{RamakrishnanArunachalam1990}.

Equation \ref{eq: por1} provides a simple model to evaluate $E_{\text{por}}$ in a solid with macroscopic pores. However, as described in Section~\ref{subsec: HP-XRD results}, lattice disorder or related phenomena at the chemical bond scale (e.g, changes in bond lengths and angles) may also affect the mechanical properties of extraterrestrial olivine. These effects may be expected to have a bearing not only on the compression behavior determined by HP-XRD, but also on the nanoindentation results. 

Thus, for the disorder porous solid, $E_\text{ref}$ in Equation \ref{eq: por1} should actually correspond to the Young's modulus of an hypothetical pore-free disordered material (in the present case, this would correspond to the extraterrestrial olivine without macroscopic pores). In order to describe the elastic modulus of extraterrestrial olivine as a function of $\varphi$, relative to the terrestrial one (i.e., non-porous, crystalline), we realize that, as a first approximation, $E_{\text{ref}} = E_0 + \Delta E$, where $E_0$ is the Young's modulus of terrestrial (pristine) olivine, and $\Delta E$ is the corresponding deviation from the terrestrial values caused uniquely by lattice disorder and/or related chemical bond effects. A similar equation can be written for the bulk modulus: $B_\text{ref}=B_0 + \Delta B$. For the sake of simplicity, we can assume that the material is isotropic and, as a first approximation, any lattice-disorder changes to $\nu$ can be neglected. Therefore, $E_\text{ref} = 3B_\text{ref}(1 -2\nu)$ and, consequently, 
\begin{equation}
    \Delta E = 3\Delta B (1 -2 \nu).
    \label{eq: delta E}
\end{equation}
In principle, it is not expected that the isothermal bulk modulus obtained by HP-XRD may be affected by the presence of macroscopic pores. This has to do with the fact that, owing to the coherence length of a few hundreds of nanometers for the diffracted radiation, XRD mainly probes the material at the scale of atomic interactions. Note, however, that nanometric deformations due to nanovoids or nanoporosity could have some influence on the HP-XRD behavior, which we neglect here. Thus, for a given olivine composition, it can be assumed that $B_0$ in the expressions above is related to high-crystallinity terrestrial olivine, while $B_\text{ref}$ corresponds to the potentially lattice-damaged extraterrestrial material. Therefore, within this assumption, we can readily evaluate $\Delta B$ from the HP-XRD data alone (see Table~\ref{tab: HP-XRD results} and Figure~\ref{fig: B_0}) and, using Equation \ref{eq: delta E}, also $\Delta E$.

Writing $E_{\text{ref}}$ in terms of $E_0$ and $\Delta E$ and substituting into Equation~\ref{eq: por1}, it is possible to obtain a phenomenological expression for the ratio $E_\text{por}/E_0$ as
\begin{equation}
    \frac{E_\text{por}}{E_0} = \left( 1 + \frac{\Delta E}{E_0} \right) \cdot \frac{\left ( 1 - \varphi \right) ^2}{1 + \kappa_E \varphi},
    \label{eq: corrected por1}
\end{equation}
with $\Delta E$ given by Equation~\ref{eq: delta E}.

\begin{figure}[t]
    \centering
    \includegraphics[width = 0.7\textwidth]{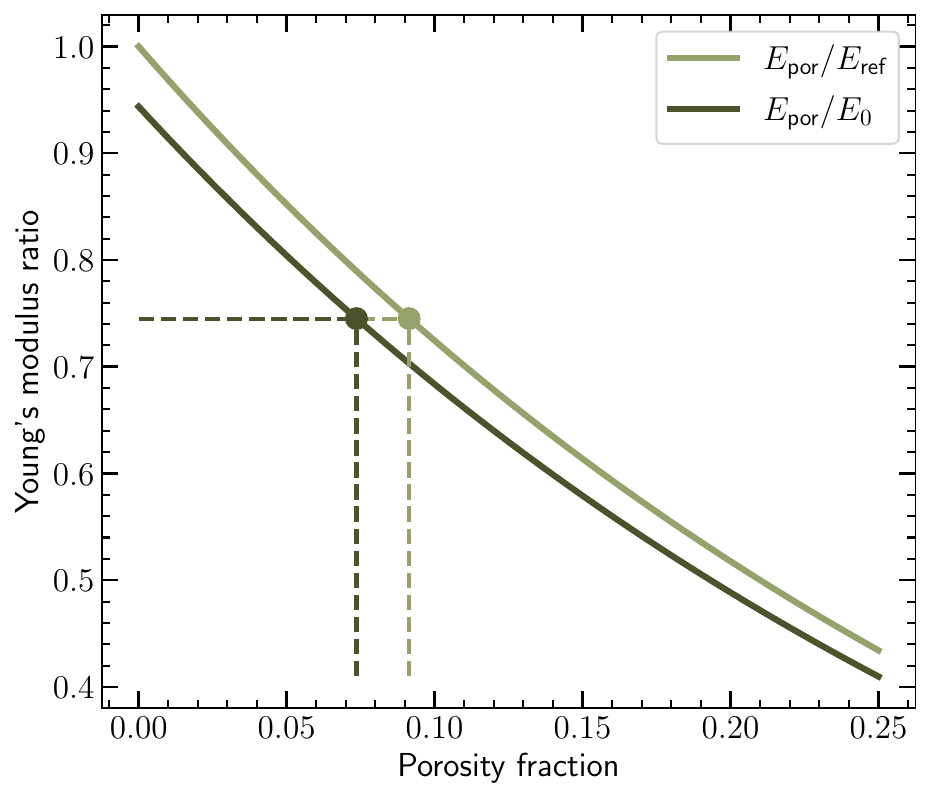}
    \caption{Young's modulus ratio between NWA~12008 olivine and reference olivines as a function of the porosity fraction of the mineral. The olive line shows the model from \citet{RamakrishnanArunachalam1990, RamakrishnanArunachalam1993} (Equation~\ref{eq: por1}). The dark olive line shows the elastic modulus ratio between porous disordered NWA~12008 olivine and non-porous crystalline terrestrial olivine (Equation~\ref{eq: corrected por1}). Scatter plots and dashed lines serve as indication, for each curve, of the required porosity fractions that reproduce the observed $E_{\text{por}} /E_0 = 0.745$ (Table~\ref{tab: nanoindentation results}).}
    \label{fig: porosity}
\end{figure}

Figure \ref{fig: porosity} shows the evolution with porosity of $E_{\text{por}}/E_{\text{ref}}$ and $E_{\text{por}}/E_0$, as obtained with Equations \ref{eq: por1} and \ref{eq: corrected por1}, respectively. For the
calculations, we used the $M_b=3B_b$ data for NWA12008 and terrestrial olivine \citep{Nestola11} together with a constant Poisson ratio of $\nu=0.273$ as given in \citet{Chung1971} for olivine with $\text{Fo}=50\%$. The figure clearly shows how, as expected, the elastic modulus of porous olivine is significantly lowered with increasing porosity in both models.

Our nanoindentation measurements on the (porous) lunar meteorite NWA~12008, together with the measurements on (reference) terrestrial olivine, allow us to obtain $E_\text{por}/E_\text{0} = 0.745$ (Table~\ref{tab: nanoindentation results}). If we first assume that this value is solely the consequence of macroscopic porosity effects then, applying Equation~\ref{eq: por1} (i.e., neglecting any lattice disorder/chemical bond effects, so that the reference material would be the terrestrial one, with $E_0$), a porosity of $\varphi = 9.0\%$ would reproduce the observed softening (Figure~\ref{fig: porosity}). 

However, as already discussed, it can be expected that both macroscopic and chemical-bond scale effects shape the Young's modulus values obtained with the nanoindentations. By using Equation~\ref{eq: corrected por1} to include such effects, the experimental $E_\text{por}/E_\text{0} = 0.745$ value for NWA~12008 yields a somewhat lower macroscopic porosity fraction, around $\varphi = 7.3\%$ (Figure~\ref{fig: porosity}). In this case, it is seen that most of the softening would still come from porosity effects, while the contribution of lattice disorder would be limited in the case of the NWA~12008 lunaite. It must be noted that in both cases the extracted porosity values are compatible with the results of porosity measurements of extraterrestrial materials \citep{Britt16_porosity, Wilkinson16_porosity, Macke10,Macke11}.

If similar calculations are performed for the Chelyabinsk meteorite, for which HP-XRD (present work) and nanoindentation data \citep{MoyanoCambero17_Chelyabinsk} are also available, the resulting porosity according to Equation~\ref{eq: por1} (neglecting lattice disorder) is $\varphi = 3.7\%$. On the other hand, using the simple model of Equation \ref{eq: corrected por1} and the experimental $M_b$ data for this ordinary chondrite (Table \ref{tab: HP-XRD results}), the corresponding macroscopic porosity extracted with Equation \ref{eq: corrected por1} is ${\varphi = 2.6\%}$. As in the case of NWA~12008, only a small part of the softening observed with the nanoindentations seems to be related to lattice disorder. In both cases it is concluded that porosity is the main responsible for such behavior. 

\section{Conclusions}\label{sec: conclusions}
In this work, we have analyzed the mechanical and high-pressure structural properties of olivine in the NWA~12008 lunaite and in some chondritic meteorites. Olivine has long been studied for its relevance in Earth sciences and, in particular, for its high abundance in Earth's mantle. In the context of extraterrestrial geochemistry, olivine is also present in high proportions in many planetary bodies, and therefore its study is relevant, among others, for ISRU or for planetary defense.

We determined the hardness and reduced elastic modulus of numerous olivine grains of the NWA~12008 meteorite using the nanoindentation technique. For completeness, we performed similar experiments on a control terrestrial olivine sample. Our results suggest that the lunar olivine is softer and more elastic by $\sim 31\%$ and $\sim 26\%$, respectively, compared with terrestrial olivine. These observations might be attributed to the interplay between two factors: macroscopic porosity (or related features) and changes at the scale of the crystal lattice. The possibility that the observed differences might be due solely to a compositional effect can mostly be ruled out, as no clear trend is seen in the results of the nanoindentations when considering the \%Fo content of the analysed olivine grains.

In order to shed light on the origin of the observed softening, we performed additional HP-XRD measurements to investigate the compressibility behavior of olivine in NWA~12008 at the crystal lattice scale. For comparison purposes, we also performed HP-XRD measurements on three different ordinary chondrites. These experiments suggest that olivine, or at least its $b$ orthorhombic axis, might be more compressible in NWA~12008 and in some chondrites relative to terrestrial olivine. These results allow us to conclude that, at least in part, the softening observed with the nanoindentations might be related to structural changes at the chemical bond scale. Lattice disorder induced by shock metamorphism or space weathering, for instance, might have affected the mechanical and elastic behavior of the mineral through a reduction of crystalline quality.

We have performed a combined analysis of the nanoindentation and HP-XRD data with a simple phenomenological model, which was aimed at disentangling the contribution of macroscopic effects (porosity) and chemical-bond scale phenomena on the mechanical properties measured with the nanoindentations. Our model suggests that the observed mechanical softening of NWA~12008 is $\sim7\%$ due to porosity and $\sim2\%$ due to lattice distortions. For the case of the higly-shocked Chelyabinsk meteorite, we infer that the mechanical properties extracted with the nanoindentations are closer to those of terrestrial olivine due to a much lower porosity.

The fact that lunar and asteroidal olivines exhibit modified mechanical behavior compared to terrestrial olivine is of great relevance both from a fundamental point of view but also for ISRU applications and future space missions. Overall, this work confirms previous results \citep{PenaAsensio24_nanoindentations,MoyanoCambero17_Chelyabinsk} suggesting that the mechanical properties of some mineral constituents of planetary materials might be different than those on Earth. Nevertheless, more work is still required in order to unambiguously confirm the present conclusions. In particular, more nanoindentations should be conducted over a larger number of olivines on other extraterrestrial bodies to confirm these findings. It would be specially interesting to measure the mechanical properties of olivine in a returned sample, other than a meteorite. With such experiments, for instance, the possible role of shock metamorphism on the properties of meteoritic materials could be discarded. 

In conclusion, the combination of nanoindentations and HP-XRD measurements to investigate the mechanical and elasticity properties of extraterrestrial minerals offers a promising avenue for future research in planetary science and may become a useful methodology to characterize planetary materials. New HP-XRD experiments ought to be performed in order to better constrain the EoS parameters of lunar and asteroidal olivine, as well as of that of other minerals from meteorites and returned samples. \\

\textbf{\large{Acknowledgements}}

Financial support from the project PID2021-128062NB-I00 funded by the Spanish Ministerio de Ciencia, Innovación y Universidades MCIU (DOI:10.13039/501100 011033) is acknowledged, as well as the Spanish program Unidad de Excelencia María de Maeztu CEX2020-001058-M. The ALBA-CELLS synchrotron is acknowledged for granting beamtime at the MSPD beamline under projects 2021095390 and 2022025734. PG-T acknowledges the financial support from the Spanish MCIU through the FPI predoctoral fellowship PRE2022-104624. JS acknowledges the financial support from projects 2021-SGR-00651 and PID2020-116844RB-C21. EP-A acknowledges financial support from the LUMIO project funded by the Agenzia Spaziale Italiana (2024-6-HH.0). DE thanks the financial support from Spanish MCIU under projects PID2022-138076NB-C41 and RED2022-134388-T from Generalitat Valenciana (GVA) through grants CIPROM/2021/075 and MFA/2022/007, which are part of the Advanced Materials program and is supported with funding from the European Union Next Generation EU (PRTR-C17.I1). RT and DE (PB and DE) thank GVA for the Postdoctoral Fellowship CIAPOS/2021/20 (CIAPOS/2023/406). JS-M thanks the Spanish MCIU for the PRE2020-092198 fellowship. NWA~12008 has been studied within the framework of an international European consortium led by IFP. Special acknowledge to I. Weber for providing the NWA~12008 meteorite thin section. This work is part of the doctoral thesis of PG-T (Doctoral Program in Physics at Universitat Autònoma de Barcelona).

\printcredits

\bibliographystyle{cas-model2-names}

\bibliography{bibliography}{}



\end{document}